\documentclass[11pt]{article}
 
\usepackage[utf8]{inputenc}
\usepackage[a4paper, total={6.5in, 8.5in}]{geometry}
\usepackage[style=phys]{biblatex}
\usepackage{graphicx}
\usepackage{authblk}
\usepackage{caption}
\usepackage{mathabx}
\usepackage[colorlinks=true,citecolor=blue,linkcolor=blue]{hyperref}

\AtEveryBibitem{\clearfield{month}}
\AtEveryBibitem{\clearfield{doi}}
\AtEveryBibitem{\clearfield{url}}
\AtEveryBibitem{\clearfield{pages}}
\AtEveryBibitem{\clearfield{issn}}

\begin{document}

\title{Carrier density crossover and quasiparticle mass enhancement in a doped 5$d$ Mott insulator}       

\author[1,2]{Yu-Te Hsu\footnote{yute.hsu@ru.nl}}
\author[3]{Andreas Rydh}
\author[1]{Maarten Berben}
\author[1]{Caitlin Duffy}
\author[4]{Alberto de la Torre}
\author[5,6]{Robin S. Perry}
\author[1,7]{Nigel E. Hussey\footnote{n.e.hussey@bristol.ac.uk}}

\affil[1]{High Field Magnet Laboratory (HFML-EMFL) and Institute for Molecules and Materials, Radboud University, Toernooiveld 7, 6525 ED Nijmegen, Netherlands}
\affil[2]{Center for Theory and Computation, National Tsing Hua University, 101, Section 2, Kuang-Fu Road, Hsinchu 300044, Taiwan}
\affil[3]{Department of Physics, Stockholm University, SE-106 91 Stockholm, Sweden}
\affil[4]{Department of Physics, Northeastern University, Boston, MA 02115, United States}
\affil[5]{London Centre for Nanotechnology and Dept. of Physics and Astronomy, University College London, London WC1E 6BT, United Kingdom}
\affil[6]{ISIS Neutron and Muon Source, Rutherford Appleton Laboratory, Harwell OX11 0QX, United Kingdom}
\affil[7]{H. H. Wills Physics Laboratory, University of Bristol, Tyndall Avenue, Bristol BS8 1TL, United Kingdom}

\renewcommand\Affilfont{\fontsize{9}{9}\itshape}

\date{}
\maketitle

\noindent
\textbf{High-temperature superconductivity in cuprates emerges upon doping the parent Mott insulator. Robust signatures of the low-doped electronic state include a Hall carrier density that initially tracks the number of doped holes and the emergence of an anisotropic pseudogap; the latter characterised by disconnected Fermi arcs, closure at a critical doping level $p^* \approx 0.19$, and, in some cases, a strongly enhanced carrier effective mass. In Sr$_2$IrO$_4$, a spin-orbit-coupled Mott insulator often regarded as a 5$d$ analogue of the cuprates, surface probes have revealed the emergence of an anisotropic pseudogap and Fermi arcs under electron doping, though neither the corresponding $p^*$ nor bulk signatures of pseudogap closing have as yet been observed. Here, we report electrical transport and specific heat measurements on Sr$_{2-x}$La$_x$IrO$_4$ over an extended doping range 0 $\leq x \leq$ 0.20. The effective carrier density $n_{\rm H}$ at low temperatures exhibits a crossover from $n_{\rm H} \approx x$ to $n_{\rm H} \approx 1+x$ near $x$ = 0.16, accompanied by a five-orders-of-magnitude increase in conductivity and a six-fold enhancement in the electronic specific heat. These striking parallels in the bulk pseudogap phenomenology, coupled with the absence of superconductivity in electron-doped Sr$_2$IrO$_4$, disfavour the pseudogap as a state of precursor pairing and thereby narrow the search for the key ingredient underpinning the formation of the superconducting condensate in doped Mott insulators.}

There are two principal pathways along which a system of itinerant electrons approaches a  Mott state \cite{imada1993}. When the interactions are predominantly local, a paramagnetic Fermi liquid survives up until the metal-insulator transition, albeit with a divergent quasiparticle mass. This type of transition is observed, for instance, in Sr-doped LaTiO$_3$ \cite{tokura1993}. The alternative transition to the Mott state is characterised by a vanishing carrier number and a non-diverging quasiparticle mass. This scenario is believed to be realised in the high-$T_{\rm c}$ cuprates \cite{ando2004, padilla2005}, though there the oxygen 2$p$ states render the system a charge-transfer insulator. Moreover, the electronic states in cuprates are $k$-space differentiated such that mobile carriers are found only near the zone diagonals, due to the presence of an anisotropic pseudogap (PG) that removes low-energy spectral weight preferentially near the zone edges and breaks the underlying Fermi surface into a series of disconnected \lq Fermi arcs' \cite{norman1998}. 

Since its discovery \cite{alloul1989}, the precise phase extent and phenomenology of the PG has been a subject of intense scrutiny and debate, though a consensus is beginning to emerge -- that of an anisotropic, states-non-conserving gap whose amplitude diminishes linearly with hole doping ($p$) and collapses at a critical doping $p^\ast$ deep within the superconducting (SC) phase boundary \cite{keimer2015, tallon2018}. Understanding the origin of PG formation is widely regarded as a key step in the elucidation not only of the Mott transition in cuprates, but also of the pairing mechanism for high-$T_{\rm c}$ superconductivity. Indeed, experimental and theoretical studies have long proposed that the PG is a signature of precursor pairing \cite{emery1995, zhou2019}. Other models incorporating non-local interactions, such as cluster dynamical mean field theory \cite{park2008} or the dual fermion approach \cite{rubtsov2008}, attribute both PG formation and electron pairing to the presence of short-range spin correlations that persist beyond the doping level at which antiferromagnetic order is destroyed. Recent Hall effect \cite{badoux2016,collignon2017} and specific heat \cite{michon2019} measurements have found evidence for an abrupt change in the carrier density ($n$) and electronic specific heat coefficient ($\gamma$) near $p^*$, prompting suggestions that the PG may terminate at a quantum critical point (QCP). Although other transport and thermodynamic properties are yet to be captured within such a scenario \cite{hussey2018}, this prospect has nonetheless motivated the discovery of a multiplicity of competing orders within the PG regime and more broadly, the search for related phenomena in layered materials that exhibit similar attributes to the cuprates. In this regard, the single-layer iridate Sr$_2$IrO$_4$ has proven to be an excellent case-study. 

Sr$_2$IrO$_4$ is isostructural to the archetypal cuprate La$_2$CuO$_4$ (albeit with a distorted array of IrO$_6$ octahedra) and features a single half-filled electronic band with $J_{\rm eff}~=~1/2$ relevant for its low-energy excitations \cite{bertinshaw2019}. The undoped compound is an antiferromagnetic Mott insulator due to the strong spin-orbit coupling of Ir 5$d$ orbitals which cooperates with the moderate Coulomb interaction to realise a Mott state. Upon $\approx 4\%$ electron doping, antiferromagnetism is destroyed \cite{chen2015} though short-range spin correlations persist to much higher doping concentrations \cite{gretarsson2016, pincini2017, horigane2018, rodan2018}. Surface-sensitive probes have revealed a PG state with striking similarities to that found in the cuprates, including a suppression of spectral weight around the zone edges and a $d$-wave energy gap by angle-resolved photoemission spectroscopy (ARPES) \cite{kim2014, kim2015, delatorre2015}, and a V-shaped PG coexisting a U-shaped Mott gap by scanning tunneling microscopy \cite{battisti2016}. As a result, Sr$_2$IrO$_4$ has been considered as a close 5$d$ analogue of the high-$T_{\rm c}$ cuprates and several theoretical studies \cite{wang2011, watanabe2013, meng2014} have predicted unconventional superconductivity under sufficient electron doping. As of now, however, superconductivity has not been confirmed. Moreover, a systematic investigation of the transport and thermodynamic characteristics of electron-doped Sr$_2$IrO$_4$, essential for understanding the nature of PG in this 5$d$ Mott insulator, has been lacking. 

Here, we report transport and specific heat measurements on a series of Sr$_{2-x}$La$_x$IrO$_4$ single crystals covering a wide range of La-doping ($0 \leq x \leq 0.20$). The evolution of the electronic ground state with carrier doping reveals striking parallels in the iridates and cuprates, which allows us to identify the critical doping level at which the PG closes in Sr$_{2-x}$La$_x$IrO$_4$. The absence of superconductivity in the iridates not only enables its electronic ground state to be studied down to low temperatures and magnetic field strengths, it also emphasises the disparate nature of the PG and SC ground states.

\subsection*{Electrical transport}
\begin{figure}[hbtp!!!]
\includegraphics[width=1\linewidth]{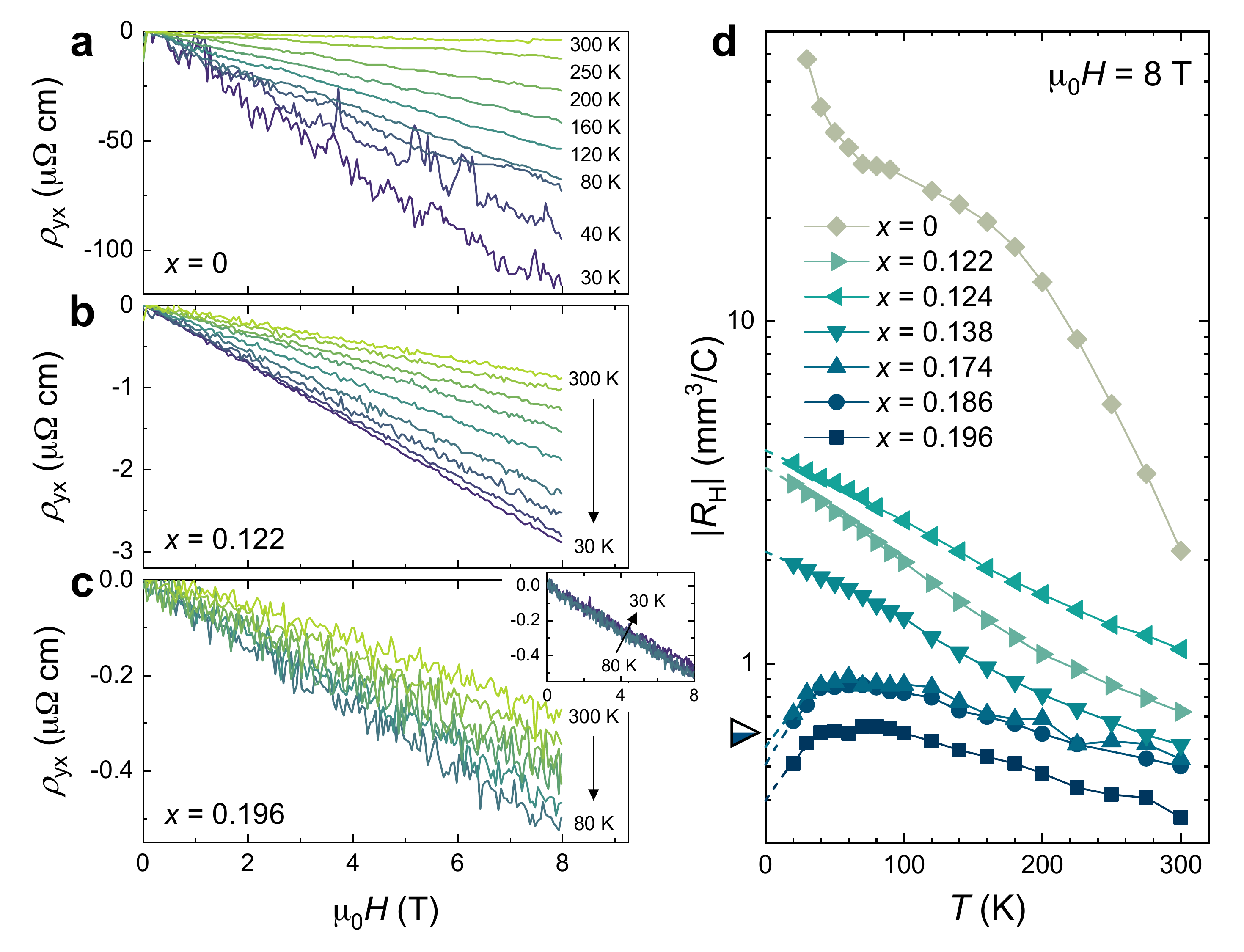}
\caption{\textbf{Hall effect in Sr$_{2-x}$La$_{x}$IrO$_4$.}
\textbf{a-c}, Magnetic-field dependent Hall resistivity $\rho_{\rm yx}(H)$ measured at specified temperatures for (\textbf{a}) $x$ = 0, (\textbf{b}) $x$ = 0.122, and (\textbf{c}) $x$ = 0.196. A negative $\rho_{\rm yx}$ that varies linearly with applied magnetic field is found for all $x$ and $T$. Note $|\rho_{\rm yx}|$ decreases by orders of magnitude as $x$ increases, reflecting a large increase in the effective Hall carrier density $|n_{\rm H}|$ induced by La doping. The high noise level for $x = 0$ at low $T$ is caused by noise in the longitudinal component of the measured signals in the highly insulating state.
\textbf{d}, Absolute Hall coefficient $|R_{\rm H}|$ as a function of La-doping $x$ and temperature $T$ plotted on a log-linear scale. Dashed lines are linear extrapolations of the $R_{\rm H}(T)$ data to the zero-temperature axis. Semi-filled triangle marks the corresponding $R_{\rm H}$ for an isotropic, fully delocalised $J_{\rm eff} = 1/2$ band at half filling.
}
\label{Hall}
\end{figure}

\begin{figure}[hbtp!!!]
\begin{centering}
\includegraphics[width=1\linewidth]{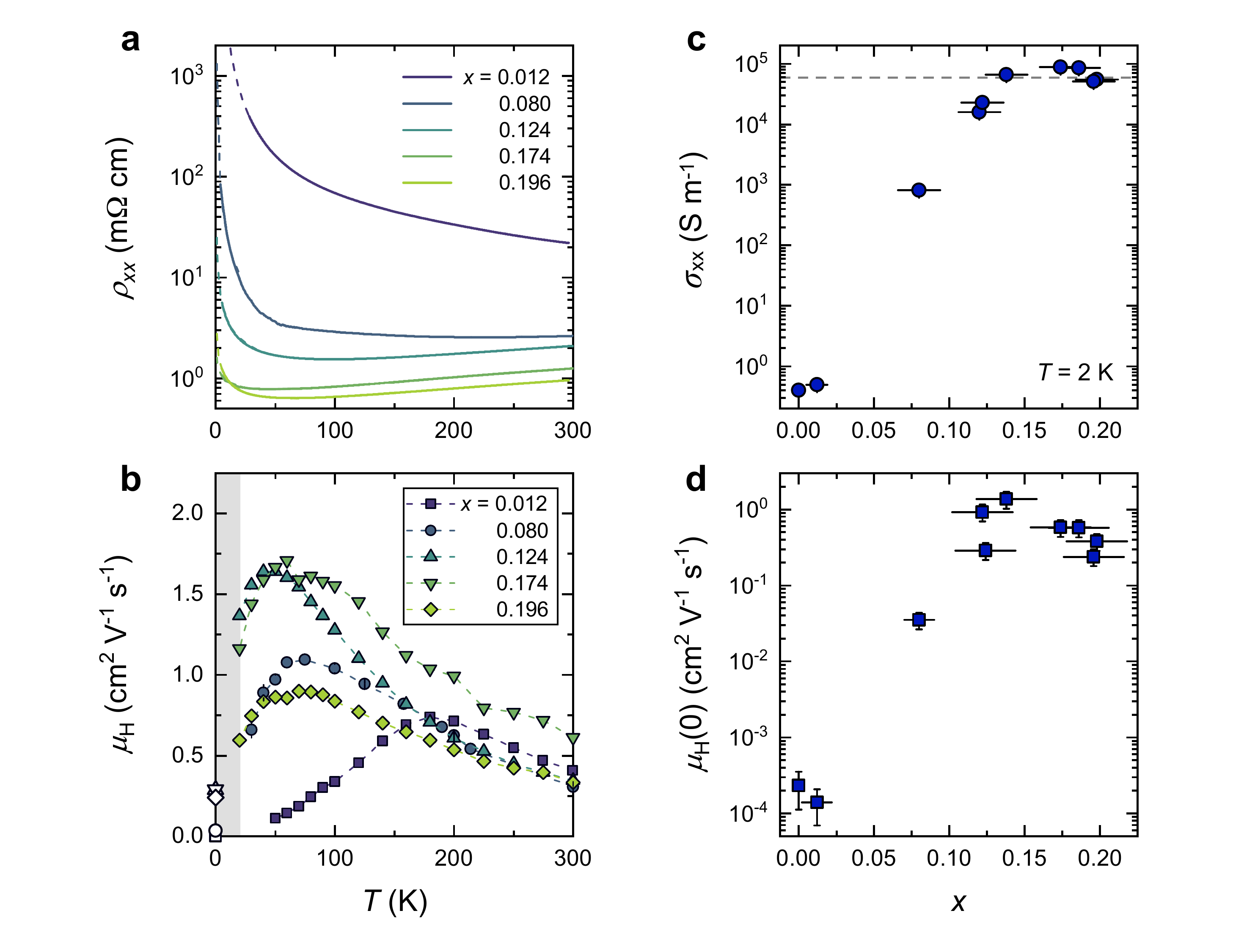}
\caption{
\textbf{Longitudinal resistivity and mobility of Sr$_{2-x}$La$_x$IrO$_4$.}
\textbf{a}, $\rho_{\rm xx}(T)$ for representative $x$ showing a rapid suppression of low-$T$ $\rho_{\rm xx}$ with electron doping. For $x = 0.035$ and $0.10$, $\rho_{\rm xx}$ data is unavailable due to the small sample size ($<400~\mu$m). Dashed lines are fits made to the data below 50~K using a variable range hopping model in three dimensions: $\rho = \rho_0$ exp$(T_0/T)^{1/4}$.
\textbf{b}, Carrier mobility calculated using $\mu_{\rm H} = R_{\rm H}/\rho_{\rm xx}$ for the same samples shown in (a). Grey shaded region is the temperature range in which $R_{\rm H}$ is not available. Open symbols at $T=0$ are calculated using the extrapolated $R_{\rm H}(T = 0)$ and fitted $\rho_{\rm xx}$ at 2~K.
\textbf{c}, Conductivity $\sigma_{\rm xx}$ at 2~K for all $x$ studied. Horizontal dashed line corresponds to the conductance quantum per IrO$_2$ plane, $\frac{h}{e^2}\frac{1}{d}$. 
\textbf{d}, $\mu_{\rm H}$ in the $T=0$ limit. Both $\sigma_{\rm xx}$(2~K) and $\mu_{\rm H}$(0) show a rapid initial increase as $x \rightarrow 0.14$, above which a moderate variation is induced by increasing $x$.
}
\label{sigma}
\end{centering}
\end{figure}

Figure~\ref{Hall} shows the Hall resistivity $\rho_{\rm yx}$ measured on an extensive series of Sr$_{2-x}$La$_x$IrO$_4$ crystals. The substitution of Sr$^{2+}$ by La$^{3+}$ introduces one additional electron per unit formula. As such, the La-content $x$ corresponds to the nominal electron doping on the Ir$^{4+}$ ion. We find $\rho_{\rm yx}$ to be negative and to vary linearly with magnetic field ($H$) even in field strengths up to 35~T (see Fig.~S1), consistent with the expectation that the doped electrons are solely responsible for the charge transport in Sr$_{2-x}$La$_x$IrO$_4$. For a material with a single type of charge carrier and an electron mean-free-path $\ell$ that is isotropic in $k$-space, the magnitude of the Hall coefficient $R_{\rm H}$ is set by the Hall carrier density $n_{\rm H}$:

\begin{equation}
    R_{\rm H} = \frac{\rho_{\rm yx}}{\mu_0 H} = \frac{1}{n_{\rm H} e},
\end{equation}
where $e$ is the elementary charge. At a fixed low temperature $T=30~$K , $|\rho_{\rm yx}|$ and $|R_{\rm H}|$ are found to decrease by more than two orders of magnitude as $x$ increases from $x$ = 0 to $x \approx$ 0.20 (Fig.~1a-c). Given the lack of any non-linearity in $\rho_{\rm yx}(H)$ that might indicate the presence of two carrier types (i.e. electrons and holes), this marked decrease in $|R_{\rm H}|$ most naturally reflects a substantial increase in the number of itinerant carriers upon electron doping. In addition, we observe a qualitative change in the temperature dependence of $R_{\rm H}(T)$ as $x$ increases beyond a threshold value $x_{\rm c} \approx 0.16$. For $x < 0.14$, $|R_{\rm H}(T)|$ increases monotonically as $T$ decreases, presumably due to the presence of an energy gap at the Fermi level $E_{\rm F}$ and a loss of itinerant carriers as thermal fluctuations are reduced. In contrast, for $x > 0.17$, $|R_{\rm H}|(T)$ exhibits a maximum at $T \approx$ 50~K, below which a modest reduction in $|R_{\rm H}|$ occurs. The marked increase in $n_{\rm H}$, coupled with the emergence of a non-monotonic form for $R_{\rm H}(T)$, suggests a recovery of the spectral weight at $E_{\rm F}$ and a marked change in the electronic ground state in Sr$_{2-x}$La$_x$IrO$_4$ beyond $x = x_{\rm c}$. At the same time, we find a sharp increase in the longitudinal conductivity $\sigma_{\rm xx}$ at low $T$ as $x \rightarrow 0.14$ (Fig.~\ref{sigma}), reaching a plateau that corresponds to the conductance quantum per IrO$_2$ plane, $\frac{h}{e^2}\frac{1}{d}$. The subsequent dip in carrier mobility $\mu_{\rm H}$ as $x$ increases beyond 0.17 likely reflects the enhanced scattering as more carriers become delocalised. Intriguingly, as later shown in Fig.~\ref{nH_gamma}, the effective zero-temperature carrier density per unit cell volume $|n_{\rm H}|V_{\rm uc}(0)$ closely tracks $n = x$ up to $x \approx 0.13$, above which $|n_{\rm H}|V_{\rm uc}$ crosses over to approaching $n = 1+x$ at $x \approx 0.20$, the maximum La-content that can be currently achieved in bulk crystals. This $n = x$ to $n = 1 + x$ evolution of the effective carrier density in electron-doped Sr$_2$IrO$_4$ across $x_{\rm c} = $ 0.16 is our first key experimental finding.

\begin{figure}[hbtp!!!]
\includegraphics[width=1\linewidth]{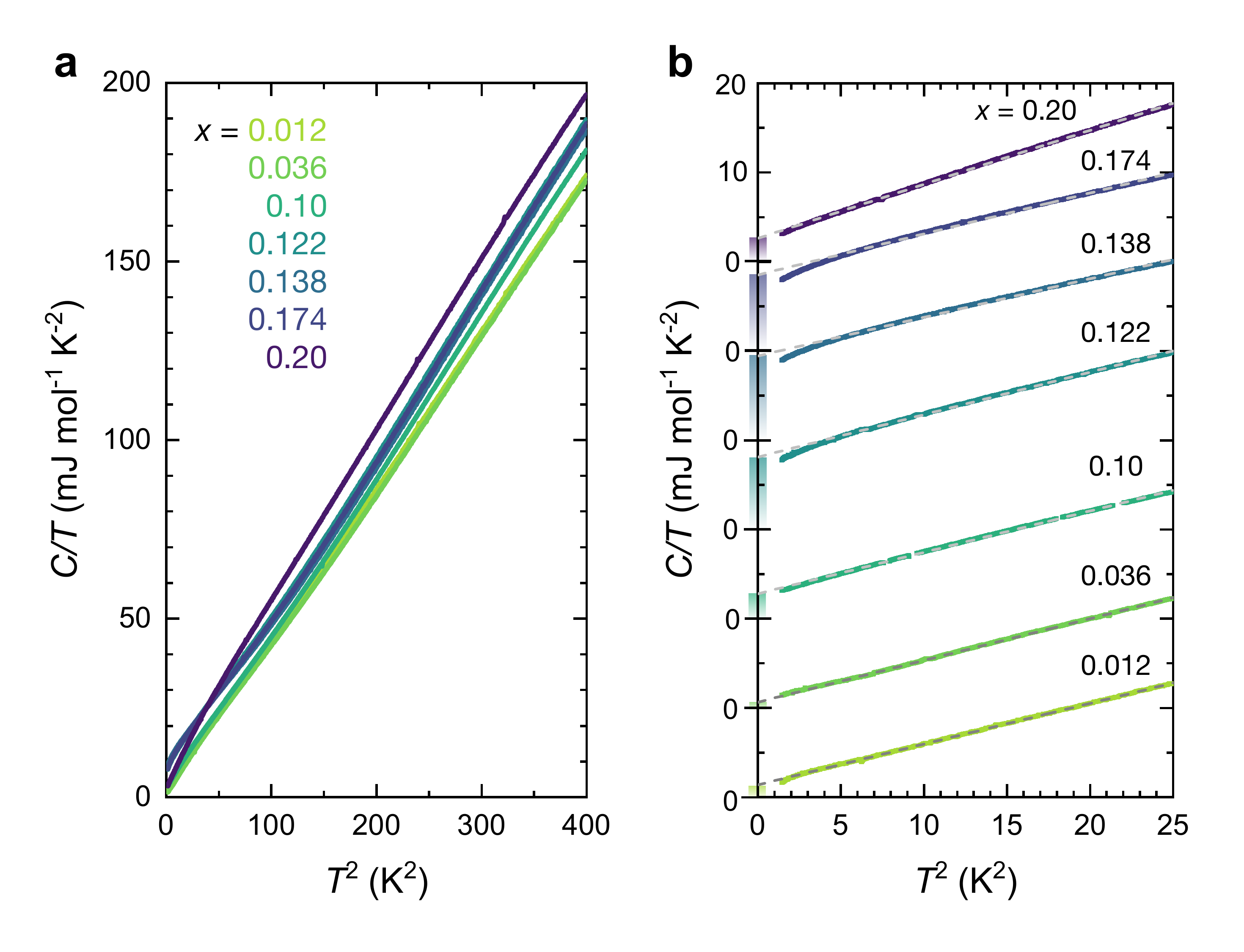}
\caption{\textbf{Low-temperature specific heat of Sr$_{2-x}$La$_x$IrO$_4$.}
\textbf{a}, Specific heat measured in zero applied magnetic field, plotted as $C/T$ versus $T^2$, between 1.5 and 20~K. A smooth, monotonic addenda background is subtracted from the data shown. The slopes $d(C/T)/d(T^2)$ between 10 and 20~K are similar for all $x$, indicating a phonon contribution that varies little with La doping. \textbf{b}, Low-temperature region ($T \leq$ 5~K) for the data shown in (\textbf{a}) with a vertical shift of 10~mJ~mol$^{-1}$~K$^{-2}$ applied to each $x$ for clarity. Grey dashed lines are fits made to the data between 2 and 5~K using $C/T = \gamma + \beta T^2$ and the vertical shadings correspond to the magnitude of zero-temperature intercepts $\gamma = C/T(T=0)$ for each $x$. 
A sizeable enhancement in $\gamma$ at intermediate La-doping $0.122 \leq x \leq 0.174$ is observed.
}
\label{Cp}
\end{figure}

\subsection*{Specific heat}
\begin{figure}[hbtp!!!]
\includegraphics[width=1\linewidth]{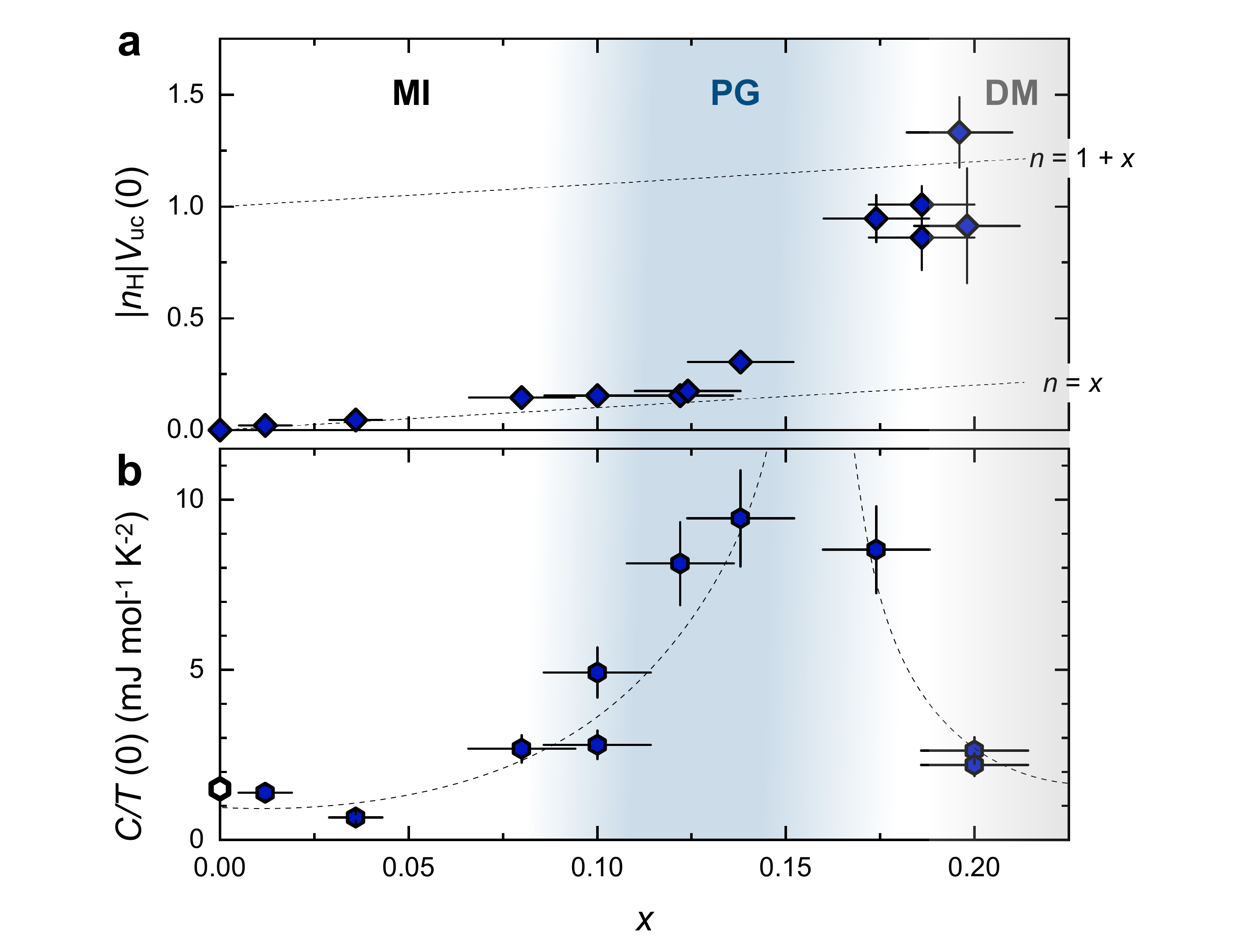}
\caption{\textbf{Doping evolution of the Hall number and specific heat in the $T=0$ limit.}
\textbf{a}, Absolute Hall number multiplied by the unit cell volume $|n_{\rm H}|V_{\rm uc}$(0), a measure of the effective carrier number per IrO$_2$ plane estimated at 0~K, as a function of $x$. Each data point is an average of the experimental $n_{\rm H}$ at the lowest measured temperature (20~K for $x > 0.06$; 30~K for $x < 0.06$) and the extrapolated $n_{\rm H}$ at 0~K (Fig.~\ref{Hall}d). The lower and upper dotted lines correspond to $n = x$ and $n = 1 + x$, respectively. Errors bars in $n_{\rm H}$ reflect the compound uncertainty in sample thickness and 0-K extrapolation (i.e. $dn_{\rm H}(0)/{n_{\rm H}(0)} = \sqrt{(dt/t)^2 + (dn_{\rm H0})/dn_{\rm H0})^2}$ where $t$ is the sample thickness and $n_{\rm H0}$ is the fitted 0-K intercept of $n_{\rm H}$). Errors in $x$ are given by the standard uncertainty of EDX measurements.
\textbf{b}, Electronic specific heat coefficient $\gamma$ estimated from the extrapolated $C/T(T = 0)$ values. Data at $x = 0$ is reproduced from previous reports \cite{carter1995, kini2006}. Dashed lines are guides to the eye. A marked enhancement in $\gamma$ is observed, centred around $x = 0.16$, coinciding with the drastic increase in $|n_{\rm H}|$. Three regimes of distinct electronic ground state are identified: a Mott insulator (MI) for $0 \leq x \lesssim 0.085$, a pseudogap state (PG) for $0.085 \lesssim x \lesssim 0.17$, and a disordered metal (DM) for $x \gtrsim 0.17$.
}
\label{nH_gamma}
\end{figure}

Measurements of the specific heat at low $T$ allow us to directly probe the electronic density of states (DOS) at $E_{\rm F}$ and track of its evolution as a function of $x$. Figure~\ref{Cp} shows the specific heat $C$ of Sr$_{2-x}$La$_x$IrO$_4$ below 20~K, plotted as $C/T$ versus $T^2$, measured on small flakes of the same crystals used for the Hall measurements. Between 10 and 20~K (i.e. 100~K$^2 \leq T^2 \leq$ 400~K$^2$), $C/T$ varies approximately linearly with $T^2$ with a slope $\beta \approx 0.43$~mJ~mol$^{-1}$~K$^{-2}$, consistent with previous reports on polycrystalline Sr$_2$IrO$_4$ \cite{carter1995, kini2006} and corresponding to a Debye temperature $\Theta_{\rm D} \approx$ 320 K. To quantitatively extract the electronic specific heat, we fit the $C/T$ data between 2 and 5~K (the lowest $T$ window in which an approximately linear-in-$T^2$ behaviour is observed) using:

\begin{equation}
    \frac{C}{T} = \gamma + \beta T^2,
\end{equation}
where $\gamma$ and $\beta$ correspond to the electronic and phononic contributions, respectively. For very lightly doped Sr$_{2-x}$La$_x$IrO$_4$ with $x$ = 0.012 and 0.036, a small $\gamma = 0.65-1.4$~mJ~mol$^{-1}$~K$^{-2}$ is found (see Fig.~\ref{Cp}b), comparable to that observed in pristine Sr$_2$IrO$_4$ ($\gamma = 1.5~\pm~0.5$~mJ~mol$^{-1}$~K$^{-2}$; ref.\cite{carter1995, kini2006}). As $x$ increases, $\gamma$ shows a substantial enhancement followed by a sharp drop at $x \approx 0.20$. The largest $\gamma$ is found for $x$ = 0.138 with $\gamma = 9.5 \pm 1.4$~mJ~mol$^{-1}$~K$^{-2}$, approximately one order of magnitude larger than found in the low-doped material. We note that the nuclear contribution is negligibly small above 1~K (see Fig.~S3). We also note a similar enhancement in the measured $C/T$ at $T=$ 2~K and 1~K (Fig.~S4), for which the phononic contribution is expected to decrease very sharply, implying that our estimate for the doping dependence of $\gamma$ is robust in the $T = 0$ limit. This large enhancement of low-$T$ specific heat centred around $x_{\rm c} = 0.16$, coinciding with the marked increase in carrier density as summarised in Fig.~\ref{nH_gamma}, is our second key finding.

For a Q2D metal, $\gamma$ is linked to the quasiparticle effective mass $m^*$ via:

\begin{equation}
    \gamma = \frac{\pi k_{\rm B}^2 a^2 N_{\rm A}}{3\hbar^2 N_{\rm uc}}\Sigma_{i} m^*_i,
\end{equation}
where $a$ is the in-plane lattice parameter, $N_{\rm A}$ the Avogadro's number, $N_{\rm uc}$ the number of formula units in a unit cell, and $i$ denotes the distinct Fermi pocket. An enhancement in $\gamma$ thus indicates a mass enhancement of the itinerant carriers, which can originate from the underlying band structure (e.g.~on approach to a van Hove singularity (vHs)) or many-body interactions (e.g.~critical fluctuations near a quantum critical point (QCP)). In quantum critical metals, the mass enhancement near a putative QCP typically manifests as a pronounced peak in $\gamma$ accompanied by a logarithmic divergence at low $T$ (i.e. $C/T~\propto$~ln(1/$T$) as $T \rightarrow 0$) \cite{loehneysen2007}. The absence of such a divergent $C/T(T)$ in Sr$_{2-x}$La$_x$IrO$_4$ (Fig.~S3), however, combined with the lack of long-range electronic order near $x$ = 0.16 (see below), appears difficult to reconcile with a scenario involving conventional quantum criticality.

\subsection*{Possible origins of the observed crossovers}
As $x \rightarrow x_{\rm c}$ from the parent Mott state, we find that the effective zero-temperature carrier density $n_{\rm H}(0)$ shows a gradual crossover from $n = x$ to $n = 1 + x$, accompanied by the conductivity per IrO$_2$ plane ($\sigma_{\rm p})$ approaching the conductance quantum and a six-fold enhancement in electronic specific heat $\gamma$. Beyond $x_{\rm c}$, $\gamma$ is found to drop markedly in a manner that is reminiscent of what is found in hole-doped La$_{1.6-x}$Nd$_{0.4}$Sr$_x$CuO$_4$ beyond $p^\ast \approx 0.23$ (ref.~\cite{michon2019, collignon2017}); likewise, the evolution of transport coefficients in Sr$_{2-x}$La$_x$IrO$_4$ is qualitatively similar to that reported in the hole-doped cuprates \cite{ando2004, laliberte2016, badoux2016, collignon2017, putzke2021}. At low doping ($x \lesssim 0.13$), $|n_{\rm H}|(x) \approx x$, implying that only the doped electrons become mobile, consistent with previous ARPES studies in lightly electron-doped Sr$_2$IrO$_4$ showing a small spectral weight at $E_{\rm F}$  \cite{delatorre2015, brouet2015}. A similar relation ($n_{\rm H}(x) \approx p$ determined at $T$ = 50~K) is found in La$_{2-x}$Sr$_x$CuO$_4$ (LSCO) below $x$ = 0.08 \cite{ando2004, badoux2016}. In YBa$_2$Cu$_3$O$_{7-\delta}$ \cite{badoux2016} and Nd-doped LSCO \cite{collignon2017}, a sharp crossover from $n_{\rm H}\approx p$ to $n_{\rm H} \approx 1 + p$ is observed in the vicinity of $p^*$, suggesting that the hole initially localised in the parent state is recovered as soon as the pseudogap vanishes. Such a drastic change in $n_{\rm H}$ may signify a Fermi surface reconstruction (FSR) driven by an electronic/structural order parameter with sufficiently long-ranged correlations. In Nd-LSCO, for example, this is likely to be spin stripe correlations \cite{ma2022}. In Sr$_{2-x}$La$_x$IrO$_4$, the only long-range electronic order known to exist is the $(\pi, \pi)$-antiferromagnetic (AFM) order, which shares the same reduced Brillouin zone boundary caused by the IrO$_6$ octahedra rotations. The resultant reconstructed Fermi surface of Sr$_{2-x}$La$_x$IrO$_4$ is expected to comprise both electron-like pockets around ($\pi$/2, $\pi$/2) nodal region and hole-like pockets around the (0, $\pi$) antinodal region \cite{delatorre2015}. Consequently, the measured Hall signals should include contributions from both Fermi pockets, which would typically give rise to a non-linearity of $\rho_{\rm yx}(H)$ at sufficiently high $H$ and a $R_{\rm H}(T=0)$ value that markedly deviates from expectations for a single conducting channel, both of which are not observed (see Fig.~S1). Moreover, the AFM order is known to vanish at a doping level ($x \leq 0.04$) far below $x_{\rm c}$, while the lattice distortion does not fundamentally alter the underlying electronic structure (except leading to a back-folded shadow band that is photon energy-dependent \cite{kim2014, delatorre2015, peng2022}). Furthermore, the structural distortion is not suppressed by La-doping \cite{horigane2018} thus cannot lead to a profound change in electronic structure across the doping series investigated. It is therefore unlikely that FSR driven by long-range order can explain the drop in $|n_{\rm H}|$ below $x_{\rm c} \approx$ 0.16 in Sr$_{2-x}$La$_x$IrO$_4$. 

Models of FSR in the absence of long-range order within the PG regime have been proposed for the cuprates \cite{yang2006, chatterjee2016, morice2017}, and numerical calculations \cite{storey2016, verret2017} based on the model proposed by Yang, Rice, and Zhang (YRZ) \cite{yang2006} show that both $n_{\rm H}$ and $\gamma$ can exhibit a marked enhancement near $p^*$ due to the appearance of the reconstructed electron-pocket. While it is not yet established whether the YRZ model is applicable to Sr$_2$IrO$_4$, evidence for the persistence of short-range AFM fluctuations up to 20\% electron doping has been reported \cite{pincini2017}. A recent dynamical mean-field study of Sr$_2$IrO$_4$\cite{moutenet2018} found that the electronic ground state evolves with electron doping through several distinct regimes: a Mott-insulating state at low dopings, a PG and a differentiated regime with suppressed spectral weight near the zone boundary at intermediate dopings, and a uniform metallic regime in which the electronic states at $E_{\rm F}$ become increasingly more coherent with increasing doping. Such a scenario is consistent with the $n \approx x$ to $n \approx 1+x$ crossover, though it is not yet clear how $\gamma$ and $m^*$ would evolve within such a model.

The PG in cuprates has also been interpreted as a spatially inhomogeneous, doping-dependent localisation gap $\Delta$ that at low doping localises exactly one hole per CuO$_2$ unit \cite{pelc2019}. In this picture, the $n_{\rm H} \approx p$ to $n_{\rm H} \approx 1+p$ crossover is a gradual one, as observed in certain cuprate families\cite{putzke2021}, due to its broad distribution in energy and real-space heterogeneity and that the holes originally localised in the parent Mott state only become fully delocalised at the edge of the SC dome where the PG is assumed to close. While this picture can account for a gradual crossover in $n_{\rm H}$ with doping, the link between the crossover and PG closing is not yet proven \cite{putzke2021, berben2022}. 

An alternative driver for the enhanced $\gamma$ in Sr$_{2-x}$La$_x$IrO$_4$ is a vHs crossing at which the Q2D Fermi surface undergoes a Lifshitz transition from electron-like to hole-like. In such a scenario, $n_{\rm H}$ (in the isotropic-$\ell$ limit) is expected to change sign. Were $\ell$ to be anisotropic, however, $n_{\rm H}$ may no longer correspond to the carrier concentration or its character \cite{ong1991, hussey2003} (see Methods). In analogy with the cuprates, the observation of Fermi arcs in the PG regime suggests a \lq nodal-antinodal dichotomy’ that manifests itself at higher doping with $k$-dependent scattering and a departure from the isotropic-$\ell$ limit. In this circumstance, one expects the regions with hole-like curvature near the antinodes to have a shorter $\ell$ than the electron-like sections close to the nodes. Consequently, the Hall conductivity $\sigma_{\rm xy}$ will continue to be dominated by those regions of the Fermi surface with electron-like curvature and $R_{\rm H}$ will remain negative. The hole-like regions will nevertheless contribute progressively to the total conductivity $\sigma_{\rm xy}$ and $\sigma_{\rm xx}$, resulting in a reduction in $|R_{\rm H}|$ and a corresponding growth in $|n_{\rm H}|$. Thus, the enhancements in $|n_{\rm H}|$, $\sigma_{\rm xx}$, and $\gamma$ may be consistent with a scenario based on a vHs crossing near $x_{\rm c} \approx$ 0.16.

While the coincidence of the vHs crossing and PG closure appears to be well-established for LSCO \cite{zhong2022}, it is not the case in Sr$_{2-x}$La$_x$IrO$_4$. Previous band structure calculations find that the vHs in Sr$_2$IrO$_4$ lies at approximately 0.12~eV above the chemical potential (irrespective of the inclusion of lattice distortions), corresponding to an electron doping exceeding 20\% \cite{watanabe2013, yang2014, lane2020}. This conclusion is confirmed by our tight-binding calculations that locates the vHs at $x \approx$ 0.29 (see Supplementary Materials). All these calculations appear to place the vHs crossing in Sr$_{2-x}$La$_x$IrO$_4$ well beyond $x_{\rm c}$, though future ARPES experiments on more highly doped crystals are required to confirm its location and clarify its impact on the thermodynamic properties. Irrespective of the vHs location, without PG closure and a contribution from the hole-like regions to $\sigma_{\rm xy}$, the measured $n_{\rm H}$ would not approach $n = 1 + x$. Moreover, the $n_{\rm H}$ crossover to $n = x$ would not occur without PG opening as the $\ell$-anisotropy required to reproduce the crossover is unphysically large (see Supplemental Materials). Hence, we can conclude that the drop in $n_{\rm H}$ from $n = 1 + x$ below $x_{\rm c}$ is associated with a loss of itinerant carriers upon PG opening. This observation not only identifies the location of PG closure, it also places Sr$_{2-x}$La$_x$IrO$_4$ alongside the cuprates in the second category of doped Mott insulators with the electronic specific-heat coefficient diminishing upon approaching the Mott state, and implies that non-local interactions must play a prominent role in the physics of the iridates.

Despite the striking similarities in the normal-state properties of Sr$_{2-x}$La$_x$IrO$_4$ and the cuprates, to date no robust signatures of superconductivity have been reported in the former. In this regard, it is instructive to compare the characteristic energy scales in Sr$_2$IrO$_4$ to those in isostructural La$_2$CuO$_4$. We summarise in Table~S2 the leading terms associated with carrier hopping ($t$), spin correlations ($J$), and charge correlations ($U$) of the two compounds. (Note for Sr$_2$IrO$_4$, $J$ characterises the Hund's coupling which promotes spin exchange \cite{watanabe2010}.) As the 5$d$ orbitals are more spatially extended, $t$ is larger in Sr$_2$IrO$_4$ while $J$ and $U$ are smaller. It has been pointed out previously\cite{watanabe2010} that the phase space available for stable superconductivity reduces with increasing $J/U$ while it expands with increasing $U/t$, with a suggested critical value of $U/t >$ 6 for the realisation of superconductivity. Though $J/U$ is similarly small in both oxides, $U/t$ is substantially reduced for Sr$_2$IrO$_4$ and very close to (or slightly below) the critical value. It is possible that a distinctly favourable energy landscape for superconductivity is only realised in the cuprates, which marks the key difference between the two oxide families, despite sharing the same pseudogap phenomenology. A more mundane reason for the absence of superconductivity is the higher level of disorder reflected in the low-$T$ resistivity upturns (at sufficiently high dopings). In underdoped LSCO ($x$ = 0.06), the in-plane resistivity undergoes a minimum \cite{ando2004} with $\rho_{ab}~\approx$~600~$\mu\Omega$~cm at 50~K. By contrast, in Sr$_{2-x}$La$_x$IrO$_4$ ($x \approx$ 0.17, i.e. with a significantly higher carrier density), $\rho_{ab} \approx$~650~$\mu\Omega$~cm at 50~K, implying a lower overall carrier mobility. The growth of cleaner iridates, were it ever to be possible, would obviously help to address this point. The PG, meanwhile, is seen to be robust to the presence of disorder.

In summary, transport and thermodynamic measurements on a series of Sr$_{2-x}$La$_x$IrO$_4$ crystals reveal two defining bulk signatures of the fate of carriers doped into a 5$d$ Mott insulator Sr$_2$IrO$_4$: the density of the itinerant carriers undergoes a dramatic reduction upon the removal of electron dopants at $x_{\rm c} \approx 0.16$ while the electronic specific-heat coefficient is strongly enhanced upon approaching $x_{\rm c}$. Our findings indicate that the low-$T$ electronic ground state of Sr$_{2-x}$La$_x$IrO$_4$ undergoes a profound transformation with electron doping and, 
aided by numerical calculations, we conclude that 1) the loss of itinerant carriers signifies the pseudogap opening near $x_{\rm c}$; 2) the concomitant enhancement in quasiparticle effective mass is not associated with an identifiable long-range order parameter or a QCP; 3) a bulk pseudogapped state does not promote superconductivity in a doped 5d Mott insulator.
The striking similarity between the PG phenomenology of the hole-doped cuprates and electron-doped Sr$_2$IrO$_4$ thus suggests that PG formation is generic to doped Q2D antiferromagnetic Mott insulators but is not of itself a precursor to superconductivity.
\clearpage

\section*{Methods}
\subsection*{Single crystal growth}
Crystals of Sr$_{2-x}$La$_x$IrO$_4$ are grown using a flux method with an off-stoichiometric mixture of SrCO$_3$:IrO$_2$:La$_2$O$_3$ and anhydrous SrCl$_2$ flux in a small Pt crucible (30~ml) with a loose lid. For $x < 0.1$, starting materials with a typical ratio of SrCO$_3$:IrO$_2$:La$_2$O$_3$:SrCl$_2$ = $1.9-2x:1.0:x/2:9.0$ are heated to 1245~$^\circ$C, held for 12 hours, and cooled to 1100~$^\circ$C at a rate $\approx$ 8~$^\circ$C/hour then quenched to room temperature. For $x > 0.1$, starting materials with a typical ratio of SrCO$_3$:IrO$_2$:La$_2$O$_3$:SrCl$_2$ = $1.5:1.0:18:9.0$ are heated to 1350~$^\circ$C and held for 30 hours then quenched. The resulting square platelet crystals have a typical lateral size of 0.2--2.0~mm and are separated from the flux by rinsing with de-ionised water. La content $x$, which corresponds to the nominal electron doping on the Ir site, is determined for each measured crystal using energy-dispersive X-ray (EDX) spectroscopy on cleaved surfaces using a JEOL JSM-6610LV SEM from Oxford Instrument EDS, with a typical uncertainty in $x$ of 0.01--0.02. It is worth noting that the reproducibility between growth runs was poor with identical starting constituents producing different results; a 50\% variation in EDX-measured $x$ values is often found across each batch compared to the nominal $x$.

\subsection*{Magnetotransport measurements}
Platelets of Sr$_{2-x}$La$_x$IrO$_4$ with a typical dimensions of $(600 \times 250 \times 75)$ $\mu$m$^3$ are selected for electrical transport measurements. Gold wires of 25 or 50~$\mu$m diameter are attached to the crystals using DuPont 6838 silver paint, with the crystal sides carefully painted to eliminate the $c$-axis contribution, and annealed in flowing oxygen at 450~$^\circ$C for 10 minutes to achieve sub-Ohm contact resistances. Measurements in magnetic fields up to 8~T, applied along the $c$-axis, are performed using a Cryogen-Free Measurement System by Cryogenic Ltd and measurements up to 35~T are performed using a Florida-Bitter magnet coupled to a $^4$He-flow cryostat at the High Field Magnet Laboratory at Radboud University. Longitudinal resistivity measurements were performed down to $\approx$ 3~K (except for $x$ = 0.012 down to 20~K) and Hall measurements down to 20~K. We note that small temperature fluctuations at $T < 20$~K, where all the samples show a resistive upturn with a sizeable $|d\rho_{\rm xx}/dT|$ (see Fig.~\ref{sigma}), lead to increased noise levels in the measured the Hall signals (due to imperfect contact alignment) and prevent a reliable extraction of the Hall resistivity below 20~K. The noise in our $\rho_{\rm yx}$ measurements is at a constant level $\sim 0.02~\mu\Omega$~cm for $x >0.10$. For $x \leq 0.10$, the noise level is highly dependent on $x$ and $T$ due to the divergent longitudinal component, and reaches 2$~\mu\Omega$~cm at 30~K for $x=0$. Nonetheless, the percentage uncertainty in $R_{\rm H}$ at 8~T is less than 5\% for all $x$ and $T$, and accounted for by the error bars of the corresponding $n_{\rm H}(0)$.

\subsection*{Specific heat measurements}
The present series of Sr$_{2-x}$La$_x$IrO$_4$ crystals has a low mass ($\lesssim$~1~mg) which limits the feasibility of standard specific heat methods. In order to perform measurements on miniature crystals, we adopt a Si-N membrane-based nanocalorimetry method which has been successfully employed on materials with sub-$\mu$g masses \cite{tagliati2012}. Samples from the same crystals used for the transport measurements were cleaved to yield flakes with clean surfaces with 50--100 $\mu$m in lateral dimensions and 10--30 $\mu$m in thickness. Measurements were performed in a Bluefors dilution refrigerator with a differential calorimeter operating using the ac steady-state method, with automatic adjustments of thermometer bias, ac heaters, and temperature oscillation frequency. Samples were mounted with a very small amount of Apiezon N grease, measured separately before sample mounting. The NiCrSiO$_x$ thin-film cermet calorimeter thermometers \cite{fortune2023} were operated at the $n^{\rm th}$ harmonic ($n$ between 3 and 8) of the ac heater, proving the temperature oscillation response at the $n+2$ harmonic. Absolute temperature control was obtained through a local offset heater on the calorimeter membrane, with the dilution refrigerator sitting at base temperature. Reference cell thermometer was used only as a thermometer bridge, providing a clean, high-resolution differential measurement of the temperature oscillation.

\subsection*{Numerical calculations}
We adopt the tight-binding model developed in ref.\cite{watanabe2013} to parameterise the Fermi surface of Sr$_2$IrO$_4$. The doping of electrons is modelled by shifting the chemical potential to match the enclosed Fermi surface area with the expected value (i.e. corresponding Luttinger count). We find the van Hove singularity (vHs) lies at $n \approx 1.29$ (i.e. 29\% electron doping), consistent with a previous report which found the vHs locates above 20\% electron doping \cite{yang2014}. For the electronic specific heat $\gamma$, we used the electronic density of states $D$ derived from the tight-binding Fermi surface parametrisation:

\begin{equation}
    \gamma = \frac{\pi^2}{3}k_{\rm B}^2 D,
\end{equation}

\begin{equation}
    D = \frac{1}{2\pi^2 d}\int^{2\pi}_0 \frac{k_{\rm F}}{\hbar v_{\rm F}}d\theta,
\end{equation}
where $k_{\rm F} (v_{\rm F})$ are the Fermi wavevector (velocity), $d$ is the distance between two IrO$_2$ planes, and $\theta$ is the azimuthal angle around the 2D Fermi surface. For longitudinal and Hall resistivities, we used the Boltzmann transport equations generalised to a 2D system with anisotropic scatterings:

\begin{equation}
    \sigma_{ij} = \frac{e^2}{2\pi^2\hbar d} \int^{2\pi}_0 \tau \frac{\mathbf{v_i} \cdot \mathbf{v_j}}{|v_{\rm F}|\cos\phi} k_{\rm F}d\theta,
\end{equation}
where $\phi$ is the angle between $v_{\rm F}$ and $k_{\rm F}$.

\section*{Data availability}
The source data used to create the main figures presented are provided with this paper. Any additional data are available from the corresponding authors upon request.

\section*{Code availability}
The codes associated with the band structure and transport coefficient calculations that support this study are available from the corresponding authors upon request.

\section*{Acknowledgements}
We thank F. Baumberger, A. Georges, and A. Tamai for fruitful discussions. We would like to thank R. D. H. Hinlopen for assistance with numerical calculations and G. Stenning and D. Nye for help on the PPMS and Smartlab  instruments in the Materials Characterisation Laboratory at the ISIS Neutron and Muon Source. We acknowledge the support of HFML-RU/NWO-I, member of the European Magnetic Field Laboratory (EMFL).  This work was supported by the European Research Council (ERC) under the European Union's Horizon 2020 research and innovation programme (grant agreement no. 835279-Catch-22), by the Swedish Research Council D. Nr. 2021-04360, and by the UK Engineering and Physical Sciences Research Council grants EP/N034694/1 and EP/V02986X/1.

\section*{Author contributions}
YTH designed the project.
RSP and AT grew and characterised the crystals.
YTH, MB, CD performed the magnetotransport measurements.
AR performed the specific heat measurements.
YTH performed the numerical calculations.
YTH and NEH wrote the manuscript with inputs from all authors.

\section*{Competing interests}
The authors declare no competing interests.

\printbibliography

\end{document}